\documentstyle[aps,12pt,epsfig,manuscript,amsbsy]{revtex}
\global\firstfigfalse
\begin{document}
\preprint{INJE-TP-99-9, hep-th/9912288}
\def\overlay#1#2{\setbox0=\hbox{#1}\setbox1=\hbox to \wd0{\hss #2\hss}#1%
\hskip -2\wd0\copy1}

\title{S-wave absorption of scalars by noncommutative D3-branes}

\author{ Y.S. Myung, Gungwon Kang, and H.W. Lee}
\address{Department of Physics, Inje University, Kimhae 621-749, Korea}

\maketitle

\begin{abstract}
On the supergravity side, we study the propagation of the RR scalar 
and the dilaton in the D3-branes with NS $B$-field.
To obtain the noncommutative effect, 
we consider the case of $B\to \infty(\theta \to\pi/2)$.
We approximate this
as the smeared D1-brane background with $F_5=H=0$.
In this background, the RR scalar induces an instability of the 
near-horizon geometry.
However, it turns out that the RR scalar is nonpropagating, 
while the dilaton is a physically propagating mode.
We calculate the s-wave absorption cross section 
of the dilaton. One finds 
$\sigma_0^\phi \vert_{B\to\infty} \sim (\tilde \omega \tilde R_{\pi \over 2} )^{8.9} / \omega^5$ 
in the leading-order while
$\sigma_0^\phi\vert_{B=0} \sim (\tilde \omega R_0)^{8}/\omega^5$
in the D3-branes without $B$-field.
This means that although the dilaton belongs to a minimally coupled 
scalar in the absence of $B$-field, it becomes a sort
of fixed scalar in the limit of $B \to \infty$.

\end{abstract}

\newpage
\section{Introduction}
\label{introduction}
Recently noncommutative geometry has attracted much interest in studying 
on string and M-theory in the 
$B$-field\cite{Con98JHEP02003,Dou98JHEP02008,She99PLB119,Has9907166,Mal9908134,Sei9908142,Big9908056}.
For simplicity, we consider supergravity solutions which 
are related to D3-branes with NS $B$ field.
According to the AdS/CFT correspondence\cite{Mal98ATMP231}, 
the near horizon geometry 
of D=7 black hole solution can describe the large $N$ limit of 
noncommutative super Yang-Mills theory (NCSYM).
We take a decoupling limit to isolate the near horizon geometry 
from the remaining one.
It turns out that the noncommutativity affects the ultra violet(UV) regime, 
leaving the infra red(IR) regime of the Yang-Mills dynamics unchanged.
The NCSYM is thus not useful for studying the theory at short distances.
It is well known that an NCSYM with the noncommutativity scale $\Delta$ 
on a torus of size $\Sigma$ is equivalent to an ordinary supersymmetric 
Yang-Mills theory (OSYM) with a magnetic flux provided that 
$\Theta = \Delta^2/\Sigma^2$ is a rational number\cite{Has9911057}.
The equivalence between the NCSYM and the OSYM can be understood
from the T-duality of the corresponding string theory.
Hence the OSYM with $B$-field is the proper description in the UV region, 
while the NCSYM takes over in the IR region. 
Actually, the noncommutativity comes from the $B\to\infty$ limit 
of the ordinary theories\cite{Mal9908134,Sei9908142,Myu9911031}.

We remind the reader that, aside the entropy, there exists an 
important dynamical quantity, ``the greybody factor(absorption cross section)'' 
for the quantum black 
hole\cite{Gub97NPB217,Gub99CMP325,Gub98NPB393,Lee98PRD104013}.
On the string side, there was a calculation for the absorption of 
scalars into the noncommutative D3-branes\cite{Hyu9909059}.
However, the authors have not considered the RR sectors in 
Ref.\cite{Hyu9909059}.

Myung, Kang, and Lee\cite{Myu9911193}have studied the quantum 
aspect of the D3-brane black hole
in $B_{23}$-field background using a minimally coupled scalar.
Such minimally coupled field might describe fluctuations of 
the off-diagonal gravitons polarized parallel to the brane 
($h_{ab}, a,b=0,1,2,3$).
They derived the exact form of the absorption cross section($\sigma_l$) 
in $B$-field on the supergravity side. 
It is well known that the cross section can be extracted from the 
solution to the linearized equation after the diagonalization.
It turns out that $\sigma_l^{B\ne0} > \sigma_l^{B=0}$.
This implies that the presence of the $B$-field suppresses the curvature effect 
surrounding the D=7 black hole.
As a result, it comes out the increase of greybody factor.

Recently, Kaya\cite{Kay9911183} has calculated the 
absorption cross section for the 
RR scalar in the D3-brane with the large $B$-field.
He claimed that the greybody factor does not change even if the $B$-field
is large.
Hence the RR scalar turns out to be a minimally coupled scalar with 
$B=0$ even for the presence of the large $B$-field.

In this paper we study the propagations of the 
RR scalar($\chi$) and the dilaton($\Phi$) by D3-branes 
with $B$-field along their world volume directions($x_2, x_3$).
Especially we are interested in the case of $B\to\infty$.
Here we use all information contained in the equations of motion,
the Bianchi identities, and the gauge condition for graviton.
In the absence of $B$-field, such fields as well as 
gravitons polarized parallel to the D3-brane belong to minimally 
coupled scalars.
However, in the presence of $B$-field,
these scalars are coupled to the background {\it nonminimally}.
In this sense, we may regard such fields as the fixed scalars.
Actually, in the smeared D1-brane background, the dilaton(RR scalar) 
turns out to be (non)propagating modes.
It is very important to test whether or not there is a 
change in the absorption cross sections of the 
dilaton between $B=0$ and $B\to\infty$ cases.

The organization of this paper is as follows.
In Sec.\ref{formalism}, we briefly review the field equations 
relevant for our study.
Here we study $B\to\infty$ limit carefully and introduce 
the smeared D1-brane black hole.
Sec.\ref{perturbations} is devoted to analyzing the perturbations 
around the smeared D1-brane background.
The propagation of the RR scalar is investigated in Sec.\ref{RR-scalar}.
This induces an instability of the near-horizon geometry.
Sec.\ref{dilaton} deals with the propagation of the dilaton with the 
dilaton gauge.
In Sec.\ref{harmonic-gauge-sec}, we study the propagation of 
the dilaton with the harmonic gauge and 
obtain its absorption cross section.
We discuss our results in Sec.\ref{discussions}.
Finally we present the smeared D1-brane solution in Appendix \ref{appendix}.

\section{Formalism}
\label{formalism}
We start with the low energy limit of type IIB superstring 
action in the Einstein frame($g_{MN} = e^{-\Phi/2} G_{MN}$)\cite{Gub97NPB217}
\begin{eqnarray}
S^E_{10} &=& { 1\over 2 \kappa_{10}^2} \int d^{10} x \Bigg [
\sqrt{-g} \Big \{ R - 
{1 \over 2} \left ( \nabla \Phi \right )^2
-{1 \over 12} e^{-\Phi} \left ( \partial B_2 \right )^2 
-{1 \over 2} e^{2 \Phi} \left ( \partial \chi \right )^2 
\nonumber \\ 
&&
-{1 \over 12} e^{\Phi} \left ( \partial C_2 - \chi \partial B_2 \right )^2 
-{ 1 \over {4 \cdot 5!}} F_5^2 \Big \}
-{1 \over {2 \cdot 4! \cdot (3!)^2 }} \epsilon_{10} 
C_4 \partial C_2 \partial B_2 \Bigg ]
\label{Einstein-action}
\end{eqnarray}
where $\Phi$ is the dilaton, $\chi$ the RR scalar, $B_2$ the NS 
two form, $C_2$ the RR two form, and $C_4$ the RR four form.
And one has
\begin{eqnarray}
&&H_{MNP} = \left ( \partial B_2 \right )_{MNP} 
   = 3 \partial_{[M} B_{NP]}, ~~
F_{3MNP} = \left ( \partial C_2 \right )_{MNP} 
   = 3 \partial_{[M} C_{2NP]},
\nonumber \\
&&\left ( \partial C_4 \right )_{MNPQR} = 
  5 \partial_{[M} C_{4NPQR]},
~~ F_5 = \partial C_4 + 5 ( B_2 \partial C_2 - C_2 \partial B_2 ).
\label{fields}
\end{eqnarray}
with the self-duality constraint $F_5 = \tilde F_5$ at the level 
of the equations of motion.
The relevant equations of motion lead to\cite{Das9911137} 
\begin{eqnarray}
&&\nabla^2 \chi + 2 \nabla \Phi \nabla \chi 
+ {e^{-\Phi} \over 6} \left ( F_3 - \chi H \right ) \cdot H = 0,
\label{eq-chi} \\
&&\nabla^2 \Phi + {1 \over 12} 
\left \{ 
e^{-\Phi} H^2 - e^{\Phi} \left ( F_3 - \chi H \right )^2 
\right \} - e^{2 \Phi} \left ( \nabla \chi \right ) ^2 =0,
\label{eq-Phi} \\
&&\nabla_M \left ( e^{-\Phi} H^{MPQ} \right ) 
- \nabla_M \left \{ \chi e^\Phi \left ( F_3 - \chi H \right )^{MPQ} \right \}
+ {2 \over 3} F^{PQRST}F_{3RST} =0,
\label{eq-FF} \\
&&\nabla_M \left \{ e^\Phi \left ( F_3 - \chi H \right )^{MPQ} \right \}
- {2 \over 3} F^{PQRST}H_{RST} =0,
\label{eq-FH} \\
&&\nabla_M F^{MPQRS} =0.
\label{eq-F}
\end{eqnarray}
In the string frame, Eqs. (\ref{eq-chi})-(\ref{eq-F}) take the following forms 
\begin{eqnarray}
&&\nabla^2_s \chi + 
 {1 \over 6} \left ( F_{3s} - \chi H_s \right ) \cdot H_s = 0,
\label{seq-chi} \\
&&\nabla^2_s \Phi - 2 \left ( \nabla_s \Phi \right )^2 + {1 \over 12} 
\left \{ 
H^2_s - e^{2\Phi} \left ( F_{3s} - \chi H_s \right )^2 
\right \} - e^{2 \Phi} \left ( \nabla_s \chi \right ) ^2 =0,
\label{seq-Phi}\\
&&\nabla_{sM} \left ( e^{-2\Phi} H_s^{MPQ} \right ) 
- \nabla_{sM} \left \{ \chi  
             \left ( F_{3s} - \chi H_s \right )^{MPQ} \right \}
+ {2 \over 3} F_s^{PQRST}F_{3sRST} =0,
\label{seq-FF} \\
&&\nabla_{sM} \left ( F_{3s} - \chi H_s \right )^{MPQ} 
- {2 \over 3} F_s^{PQRST}H_{sRST} =0,
\label{seq-FH} \\
&&\nabla_{sM} F_s^{MPQRS} =0.
\label{seq-F}
\end{eqnarray}
In addition, we need the remaining Maxwell equations, 
as three Bianchi identities
\begin{equation}
\partial_{[M} H_{NPQ]} = \partial_{[M} F_{NPQ]} = \partial_{[M} F_{NPQRS]} =0.
\label{bianchi}
\end{equation}

The solution of D=7 extremal black hole for the D3-branes 
with non-zero $B_{23}$-field is given as follows 
in the D=10 string frame\cite{Mal9908134}
\begin{eqnarray}
&&ds_{s}^2 = f^{- {1 \over 2}} \left \{ -dx_0^2 + dx_1^2 
      + h ( dx_2^2 + dx_3^2 ) \right \} 
     + f^{ {1 \over 2}} \left ( dr^2 + r^2 d\Omega_5^2 \right ),
   \label{metric-str}  \\
&&f = 1 + {{R^4_\theta} \over r^4 }, 
     ~~ h^{-1} = \sin^2 \theta f^{-1} + \cos^2 \theta, 
     \nonumber \\
&&\bar B_{s23} = \tan \theta f^{-1} h, ~~ e^{2 \bar \Phi} = g^2 h, \nonumber \\
&&\bar F_{s01r} = {1 \over g} \sin \theta \partial_r(f^{-1}),
    ~~ \bar F_{s0123r} = {1 \over g} \cos \theta h \partial_r(f^{-1}).
   \nonumber
\end{eqnarray}
From now on we work in the string frame and thus neglect 
the subscript ``s''.
Here the asymptotic value of $B$-field is $\bar B_{23}^\infty = \tan \theta$ and
the parameter $R_\theta$ is defined by 
$\cos \theta R^4_\theta =R_0^4(= 4 \pi g N \alpha'^2)$. 
$N$ is the number of the D3-branes and $g=g_\infty$ is the string coupling 
constant.
It is obvious that for $\theta=0(h=1)$ one recovers the ordinary D3-brane 
black hole with the standard AdS$_5\times$S$^5$ geometry in the near horizon. 
In this case we have ${\bar F}_{0123r} = {1 \over g} \partial_r(f^{-1})$,
its dual ($\tilde{\bar F}_5$), and $e^{2 \bar \Phi} = g^2$.

For $\theta\to\pi/2(h\to f)$, however, one finds the D3-brane black hole in 
the very large $B$-field and thus the effect of 
noncommutativity appears.
Here one finds a deviation from AdS$_5\times$S$^5$ in the 
near horizon.
However, it is known that in order to make connection to 
noncommutative geometry, $\theta\to{\pi\over 2}(B\to\infty)$ limit 
must be carefully taken.
In addition, we need a double scaling limit of 
$g N \to \infty, \omega^4 \alpha'^2 \to 0$ to keep the expansion 
parameter $ \omega^4 R_0^4$ very small 
in the calculation of the absorption cross section.
This implies the decoupling limit of $g\to 0, \alpha' \to 0 : g N \gg 1$ 
and the low-energy limit($\omega\to 0$).
Here we wish to take into account all of these limits only by taking 
$\alpha' \to 0$ :
\begin{equation}
\tan\theta = {{\tilde b} \over \alpha'}, ~~
g = \alpha' \tilde g , ~~
N = {{\tilde N} \over \alpha'^2},
\label{limit}
\end{equation}
where $\tilde b, \tilde g, \tilde N$ stay fixed\cite{Mal9908134}.
Further this implies that
\begin{equation}
\lim_{\theta\to{\pi\over 2}} \left ( R_\theta^4 = {R_0^4 \over \cos\theta}
  \right ) = 4 \pi \tilde g \tilde b \tilde N \equiv R_{\pi\over 2}^4.
\label{rtheta}
\end{equation}
This means that $\lim_{\theta\to{\pi\over 2}} R_\theta^4$ is nearly 
independent of $\theta$ and is finite with $R_{\pi\over 2}^4 \gg R_0^4$.
But we must choose the low-energy limit($\omega\to0$) to 
keep $\omega^4 R_{\pi\over 2}^4$ small.
Under this condition, one finds
\begin{eqnarray}
{\bar H}_{r23} &=& \tan\theta \partial_r(f^{-1} h )
~~~ \raise1.5ex\hbox{$\scriptstyle{\theta\to \pi/2}$} 
\hspace*{-30pt}
\raise.530ex\vbox{\hrule width20pt}\hspace*{-5pt}\longrightarrow 
\hspace*{-35pt}\lower1.5ex\hbox{$\scriptstyle{\alpha'\to 0}$} 
~~~~\alpha' \to 0,
\label{hr23} \\
{\bar F}_{01r} &=& {1 \over g} \sin\theta \partial_r(f^{-1} )
~~~ \raise1.5ex\hbox{$\scriptstyle{\theta\to \pi/2}$} 
\hspace*{-30pt}
\raise.530ex\vbox{\hrule width20pt}\hspace*{-5pt}\longrightarrow 
\hspace*{-35pt}\lower1.5ex\hbox{$\scriptstyle{\alpha'\to 0}$} 
~~~~{1 \over \alpha'} \to \infty,
\label{f01r} \\
{\bar F}_{0123r} &=& {1 \over g} \cos\theta \partial_r(f^{-1} )
~~~ \raise1.5ex\hbox{$\scriptstyle{\theta\to \pi/2}$} 
\hspace*{-30pt}
\raise.530ex\vbox{\hrule width20pt}\hspace*{-5pt}\longrightarrow 
\hspace*{-35pt}\lower1.5ex\hbox{$\scriptstyle{\alpha'\to 0}$} 
~~~~ \longrightarrow {\rm finite}.
\label{f0123r} 
\end{eqnarray}
Here one obtains a sequence of 
${\bar F}_{01r} \gg {\bar F}_{0123r} \gg {\bar H}_{r23}$ in this limit.
As a result, although the flux of the RR five-form(${\bar F}_5$) counts the 
rank of the noncommutative gauge group, this is very small in 
comparison with the RR three-form(${\bar F}_3$) in the limit of 
$\theta\to{\pi\over 2}(B\to\infty)$.
Hence we can neglect the effect of ${\bar F}_5$ and ${\bar H}$ on the 
absorption cross section in favor of ${\bar F}_3$.

In the case of ${\bar F}_5 = {\bar H} =0$, 
$e^{2 \bar \Phi} = g^2 f$, ${\bar F}_{01r}={1\over g} \partial_r(f^{-1} )$, 
one obtains the smeared D1-brane solution in Appendix \ref{appendix}.
We may regard this solution as the simple one to include the 
noncommutative effect through $\tilde R_{\pi\over 2}^4 \gg R_0^4$ in 
$\tilde f = 1 + \tilde R_{\pi\over 2}^4 / r^4$.
Here $\tilde R_{\pi\over 2}^4 = (1 + \epsilon)R_{\pi\over 2}^4$ with 
$\epsilon = k^2/\tilde\omega^2=k^2/\omega^2(1 - k^2/\omega^2)^{-1}$.

\section{Perturbations}
\label{perturbations}
Now let us introduce the perturbations to derive the greybody 
factor as\cite{Lee99JHEP10014}
\begin{eqnarray}
G_{MN} &=& \bar G_{MN} + h_{MN}
\label{lin-G} \\
\chi &=& \ 0 + \eta, 
\label{lin-chi} \\
\Phi &=& \bar \Phi + \phi ,
\label{lin-Phi} \\
F_{01r} &=& \bar F_{01r} + f_{01r} = \bar F_{01r}(1+f_3) ,
\label{lin_F3} \\
H_{r23} &=& \bar H_{r23} + h_{r23} = \bar H_{r23}(1+h_3) ,
\label{lin_H3} \\
F_{0123r} &=& \bar F_{0123r} + f_{0123r} = \bar F_{0123r}(1+f_5)
\label{lin_F5}
\end{eqnarray}
with all other perturbations to be zero.
For the perturbations above, we keep the 
background symmetry up to the linearized level.
General fluctuations give a complicated 
system of differential equations:
\begin{eqnarray}
&&\nabla^2 \eta - { 1\over 6} \bar H^2 \eta =0,
\label{eq-eta} \\
&&-h^{MN} \nabla_M \nabla_N \bar \Phi 
-\bar G^{MN} \delta \Gamma^P_{MN} \nabla_P \bar \Phi 
+ \nabla^2 \phi 
- 4 \nabla \bar \Phi \cdot \nabla \phi 
\nonumber \\
&&~~~~
+ 2 \nabla_M \bar \Phi \nabla_N \bar \Phi h^{MN} 
+ {1 \over 12} \left ( 2\bar H^2 h_3 - 
            3 \bar H_{MNQ}\bar H^{PNQ} h^M_{~P} \right ) 
\nonumber \\
&&~~~~
- {e^{2\bar\Phi} \over 12} \left \{ 2 \bar F_3^2 ( \phi + f_3 ) 
     -3 \bar F_{MNQ} \bar F^{PNQ} h^M_{~P} \right \} =0,
\label{eq-hmn} \\
&&e^{-2 \bar\Phi} \left \{ (\nabla_M - 2 \nabla_M \bar \Phi ) 
         \left ( \bar H^{MNP} h_3 \right ) 
   - \left (\nabla_M h_Q^{~N}\right ) \bar H^{MQP} 
   + \left (\nabla_M h^P_{~Q}\right ) \bar H^{MQN} \right .
\nonumber \\
&&~~~~\left .
   - \left ( \nabla_M \hat h^M_{~Q}\right ) \bar H^{QNP} 
   - h^M_{~Q} \nabla_M \bar H^{QNP}
   - 2(\nabla_M \phi) \bar H^{MNP} \right \}
\nonumber \\
&&~~~~
   - 2 \nabla_M \left ( e^{-2 \bar \Phi} \bar H^{MNP} \right ) \phi
   - \nabla_M (\bar F^{MPQ} \eta ) 
   + {2 \over 3} \bar F^{PQRST} \bar F_{RST} (f_3 + f_5 ) =0,
\label{eq-hmnp} \\
&&\nabla_M \left (\bar F^{MNP} f_3 \right ) 
  - \left (\nabla_M h_Q^{~N}\right ) \bar F^{MQP} 
  + \left ( \nabla_M h^P_{~Q}\right ) \bar F^{MQN}
  - h^M_{~Q} \nabla_M \bar F^{QNP} 
\nonumber \\
&&~~~~~~~~
  - \left (\nabla_M \hat h^M_{~Q}\right ) \bar F^{QNP}
  - \nabla_M ( \bar H^{MPQ} \eta )
  - {2 \over 3} \bar F^{PQRST} \bar H_{RST} ( h_3 + f_5 ) =0,
\label{eq-fmnp} \\
&&\nabla_M \left (\bar F^{MNPQR} f_5 \right ) 
  - 4 \bar F^{MT[PQR} \nabla_M h_T^{~N]}
  - (\nabla_M \hat h^M_{~T} ) \bar F^{TNPQR} 
  - h^M_{~T} \nabla_M \bar F^{TNPQR} =0 
\label{eq-fmnpqr}
\end{eqnarray}
with $\delta \Gamma^P_{MN} = {1 \over 2} \bar G^{PQ} (
\nabla_M h_{NQ} + \nabla_N h_{MQ} - \nabla_Q h_{MN} )$.
Here we have a relation of $\bar G^{MN} \delta \Gamma^P_{MN} =
\nabla_M \hat h^{MP}$, $\hat h^{MP} = h^{MP} -{1 \over 2} \bar G^{MP} h$ with
$h= h^T_{~T}$.
Let us check the order of $g$ in each equation.
To obtain all consistent linearized equations, we have to scale $\eta$ 
in Eqs.(\ref{eq-eta}), (\ref{eq-hmnp}) and (\ref{eq-fmnp}) as $\eta / g$.
Furthermore we find from three Bianchi identities 
in Eq.(\ref{bianchi}) with Eqs.(\ref{lin_F3})-(\ref{lin_F5}) that
\begin{eqnarray}
f_3, f_5 &\to& {\rm propagating ~~ modes,} 
\nonumber \\
h_3 &\to& {\rm nonpropagating ~~ mode}.
\nonumber
\end{eqnarray}
This means that the NS $B$-field is considered as a tool for 
giving the noncommutative 
effect but it does not belong to the physically propagating field.
For the graviton modes, we may use either 
the dilaton gauge\cite{Lee98PRD7361}
\begin{equation}
\nabla_M \hat h^{MP} = h^{MN} \Gamma^P_{MN} 
\label{dilaton-gauge}
\end{equation}
or the harmonic gauge\cite{Gre93PRL2837}
\begin{equation}
\nabla_M \hat h^{MP} = 0. 
\label{harmonic-gauge}
\end{equation}
Although a choice of gauge condition
does not eliminate all of the gauge freedom, 
it simplifies the perturbation equations.
We note here that, although the equation (\ref{eq-eta}) is 
a decoupled one for the RR scalar($\eta$),
Eqs.(\ref{eq-hmnp}) and (\ref{eq-fmnp}) contain information for $\eta$.
Kaya considered Eq.(\ref{eq-eta}) only in ref.\cite{Kay9911183}.
As can be seen the dilaton equation takes a very complicate form 
coupled with various other fields. 
To decouple $\phi$ from the remaining fields, we have to do some further work.
Hence we separate the RR scalar from the dilaton.
Let us first investigate the RR scalar.  

\section{RR Scalar Propagation}
\label{RR-scalar}
Because of the RR scalar equation (\ref{eq-eta}) 
is completely decoupled from others, 
we start with an arbitrary $B(\theta)$.
A way to obtain the noncommutative effect is to include 
the momentum dependence along the world volume directions\cite{Mal9908134}.
This is because the $B_{23}$-field is set up along these directions.
Hence $x_2, x_3$ become noncommuting coordinates.
Now let us consider the spacetime dependence
\begin{eqnarray}
\eta(t,x_1, x_2, x_3, r, \theta_i ) = e^{-i \omega t}   
     e^{i(k_1 x_1 + k_2 x_2 + k_3 x_3 )}
     Y_l(\theta_1, \theta_2, \cdots, \theta_5) \eta^l(r)   
\label{defeta} 
\end{eqnarray}
with $\bar \nabla^2_{\theta_i} Y_l(\theta_i) = - l(l+4)Y_l(\theta_i)$.
$Y_l(\theta_i)$ denotes spherical harmonics on S$^5$ with unit radius.
Here $\eta^l (r)$ is the radial part of the 
$l$-partial wave of energy $\omega$.
Then Eq. (\ref{eq-eta}) takes the form
\begin{equation}
\left \{ 
{ \partial^2 \over {\partial r^2}} + 
{ 5 \over r} {\partial \over {\partial r}}
+ { h' \over h} {\partial \over {\partial r}}
  - {{l(l+4)} \over r^2 } + (\omega^2 - k_1^2 ) f 
  - {{(k_2^2 + k_3^2 ) f} \over h } 
  - {{ f'^2 \sin^2 \theta \cos^2 \theta h^2 } \over f^3 } 
\right \} \eta^l =0 
\label{etalequation}
\end{equation}
with $f' = {d \over {  dr }}  f$.
If $k_1 = k_2=k_3=0$, this is exactly the equation that 
Kaya has considered in the first version of ref.\cite{Kay9911183}.

If $\theta=0$($B$-field is turned off) and $l=0$, one finds that 
Eq.(\ref{etalequation}) reduces to the 
s-wave minimally coupled scalar($\varphi$) equation in the D=7 
black hole background\cite{Gub97NPB217}
\begin{equation}
\left \{ 
{ \partial^2 \over {\partial r^2}} + { 5 \over r} {\partial \over {\partial r}}
  + \tilde \omega^2 
  \left ( 1 + {R^4_0 \over r^4} \right )  
\right \} \varphi^0 =0 
\label{lB0equation}
\end{equation}
with $\tilde \omega = \sqrt{\omega^2 - k_1^2 - k_2^2 - k_3^2 }$
$\simeq \omega ( 1 - {k^2 \over {2 \omega^2} } )$,
$k^2= k_1^2 + k_2^2 +k_3^2$, $\omega^2 > k^2$.
The s-wave absorption cross section 
for Eq. (\ref{lB0equation}) can be obtained from the 
solution to Mathieu's equation as\cite{Gub99CMP325}
\begin{equation}
\sigma_0^\varphi \Big \vert_{B=0} = 
      {{\pi^4 (\tilde \omega R_0)^8 } \over { 8 \omega^5 } }
\label{sigmaB0l0}
\end{equation}
in the leading-order calculation.
We note here that 
$\sigma_0^\eta\vert_{B=0} = \sigma_0^\phi\vert_{B=0} = \sigma_0^\varphi\vert_{B=0}$, 
because fluctuations of both RR scalar and the dilaton 
fields belong to minimally coupled scalars when $B$-field is absent.
For an arbitrary $B$, the corresponding equations
for a minimally coupled field $\varphi$ is given by 
\begin{equation}
\left \{ 
{ \partial^2 \over {\partial r^2}} + { 5 \over r} {\partial \over {\partial r}}
  + \tilde \omega^2 \left ( 1 +  
  { {\tilde R^4_\theta} \over r^4} \right )
\right \} \varphi^0_B =0, 
\label{Bequation}
\end{equation}
where $\tilde R^4_\theta = \left ( 1 + \epsilon \right ) R^4_\theta $ with 
$\epsilon(\theta) = {k^2 \over {\tilde \omega^2 }} \sin^2 \theta < 1$. 
In the limit of $\theta\to{\pi\over 2}$, one finds 
$\tilde R_{\pi\over 2}^4 = (1 + k^2/\tilde\omega^2 ) R_{\pi\over 2}^4$.
The above equation is exactly the same form 
as in Eq.(\ref{lB0equation}) with different ``$R$''.
Thus the absorption cross section can be read off from 
(\ref{sigmaB0l0}) simply by substituting $R_0$ with 
$\tilde R_\theta$\cite{Myu9911193}
\begin{equation}
\sigma_0^\varphi \Big \vert_B = 
\sigma_0^\varphi \Big \vert_{B=0}(R_0 \to \tilde R_\theta )
= {{ \pi^4 (\tilde \omega \tilde R_\theta )^8 } \over { 8 \omega^5}}.
\label{sigmaB}
\end{equation}
For an arbitrary $B$-field, 
one always finds 
$\sigma_0^\varphi\vert_{B\ne 0} > \sigma_0^\varphi\vert_{B=0}$ because of 
$\tilde R_\theta > R_0$.

In order to transform Eq. (\ref{etalequation}) into 
the familiar equation like Eq. (\ref{Bequation}), we redefine $\eta^0$ as 
$\eta^0 = h^{-1/2} \hat \eta $. 
Then this leads to
\begin{equation}
\left \{
{ \partial^2 \over {\partial r^2}} +
{ 5 \over r} {\partial \over {\partial r}}
+ \tilde \omega^2 \left ( 1+ {{\tilde R_\theta^4 }\over r^4} \right ) 
  + {{ 4 \sin^4 \theta R_\theta^8 h^2 } \over {r^{10} f^4 } }
\right \} \hat \eta =0 ,
\label{hateta}
\end{equation}
We can rewrite the last term in (\ref{hateta}) 
in terms of $\tilde R_\theta$, 
$\tilde f = 1 + \tilde R_\theta^4 / r^4 $, 
$\tilde h^{-1} = \sin^2 \theta \tilde f^{-1} + \cos^2 \theta$ as
\begin{equation}
{{\sin^4 \theta R_\theta^8 h^2 } \over {r^{10} f^4 }} =
{{\sin^4 \theta \tilde R_\theta^8 \tilde h^2 } \over {r^{10} \tilde f^4 }} 
\left \{
1 -2 \epsilon \left (
1 - { \tilde R_\theta^4 \over {r^4 + \tilde R_\theta^4}}
-{{\tilde R_\theta^4 \cos^2\theta }\over {r^4+\tilde R_\theta^4 \cos^2 \theta}}
\right ) + {\cal O}(\epsilon^2)
\right \}.
\label{rtilde}
\end{equation}
For the leading-order calculation, it is sufficient to keep the 
first term of the RHS of Eq.(\ref{rtilde}) only.
Using $\hat \eta = r^{-5/2} \hat{\hat\eta}$, Eq. (\ref{hateta}) leads to
the Schr\"odinger-like equation as
\begin{equation}
\left (
{ \partial^2 \over {\partial r^2}} 
+ \tilde \omega^2  - \tilde V_\theta
\right ) \hat {\hat\eta} =0 ,
\label{hateta-V}
\end{equation}
where
\begin{equation}
\tilde V_\theta =
 - \tilde \omega^2 \left ( \tilde f -1 \right ) 
 + { 15 \over { 4 r^2}} 
  - {{ 4 \sin^4 \theta R^8_\theta {h}^2 } \over {r^{10} f^4 } } .
\label{tildeV}
\end{equation}
As will be shown in Eq.(\ref{app-dil-eq}), the 
first term in Eq.(\ref{tildeV}) plays a role of 
energy term with $E=1$ in the near 
horizon of $r < R_\theta$.
For $r > R_\theta$, the first term can be ignored.
Thus we can approximate $\tilde V_\theta$ as $V_\theta$
\begin{eqnarray}
V_\theta &=& {15 \over { 4 r^2}} - {{ 4 \sin^4 \theta R^8_\theta {h}^2} 
           \over {r^{10} f^4}} .
\label{Veta}
\end{eqnarray}
\begin{figure}
\epsfig{file=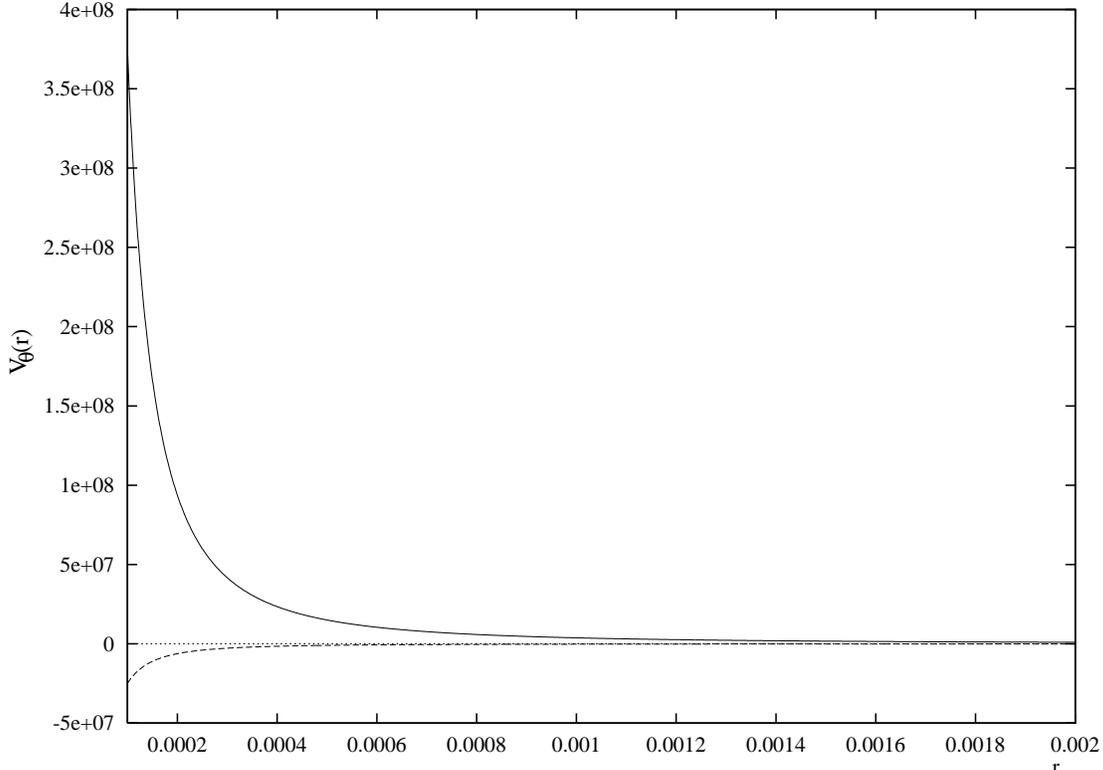,width=0.9\textwidth,clip=}
\caption{
\label{rr-scalar}
The graphs of the RR scalar potential in the near horizon.
For $\theta\to\pi/2$, one finds a potential well(dashed line) 
and for $\theta=0$, one finds a potential barrier(solid line).
The horizon is located at $r=0$.
}
\end{figure}
\noindent
For an arbitrary $\theta(B)$, it is very difficult to solve 
Eq.(\ref{hateta-V}).
Thus, let us discuss two interesting cases.
If $\theta \simeq 0 $, $h \simeq 1$.
In this case, the last term of Eq.(\ref{Veta}) can be neglected,
compared with the first one.
Then the RR scalar cross section takes the same form as that of the
minimally coupled scalar in Eq. (\ref{Bequation}).
For $\theta \to {\pi\over 2}$, $h \to f$.
In this case,  the last term of Eq.(\ref{Veta}) plays an important 
role in the near-horizon.  
In the near horizon, one finds that 
$V_{\theta=0}^{NH}=15/4r^2$ for $\theta=0$ and 
$V_{\theta\to{\pi\over2}}^{NH} = -1 /4r^2$ for $\theta\to\pi/2$.
The latter case induces an instability of the near-horizon 
geometry in the smeared D1-brane background
because the potential well allows us the scattering state($\omega=$real) 
as well as the exponentially growing state($\omega=i\Omega$).
Also the same situation is recovered if one uses $\nabla^2 \eta=0$ 
instead of Eq.(\ref{eq-eta}).
Hence the instability appears even for ${\bar H}_{MNP}=0$.
As is shown in Fig.\ref{rr-scalar}, the singular behaviors of 
$V_{\theta=0,{\pi\over2}}$ seem to appear as $r\to 0$.
However, this is a coordinate artifact.
Using the coordinate $z$ in Sec.\ref{harmonic-gauge-sec}, instead of $r$, 
one cannot find such singular behaviors in the near horizon.

\section{Dilaton Propagation with the dilaton gauge}
\label{dilaton}
In this section, we wish to study the propagation of the 
dilaton with the dilaton gauge in (\ref{dilaton-gauge}).
Under this gauge the dilaton equation takes a rather simple 
form than the harmonic gauge\cite{Lee98PRD7361}.
Assuming
\begin{eqnarray}
\phi(t,x_1, x_2, x_3, r, \theta_i ) = e^{-i \omega t} 
     e^{i(k_1 x_1 + k_2 x_2 + k_3 x_3 )}
     Y_l(\theta_1, \theta_2, \cdots, \theta_5) \phi^l(r) ,
\label{defphi}
\end{eqnarray}
the dilaton equation (\ref{eq-hmn}) leads to
\begin{eqnarray}
&&\left \{
{ \partial^2 \over {\partial r^2}} + 
{ 5 \over r} {\partial \over {\partial r}}
- { h' \over h} {\partial \over {\partial r}}
  - {{l(l+4)} \over r^2} 
  +(\omega^2 - k_1^2 ) f 
  - {{(k_2^2 + k_3^2 ) f} \over h } \right .
\nonumber \\
&&~~~~~~~~~~~~~~~~~~~~~~~~~~~~~~~~~~~~~~~~~~~~~~~~~~~~~ \left .
  + {{ 16 \sin^2 \theta \cos^2 \theta R_\theta^8 h^2 } \over {r^{10} f^3} } 
  + {{ 16 \sin^4 \theta R_\theta^8 h^2 } \over {r^{10} f^4} } 
\right \} \phi^l 
\nonumber \\
&&~~~~~~~~
+ {{ 16 \sin^2 \theta \cos^2 \theta R_\theta^8 h^2 } \over {r^{10} f^3 }} 
  \left \{ f_3 - {1 \over 2} ( h^0_{~0} + h^1_{~1} +h^2_{~2} +h^3_{~3} ) \right \}
\nonumber \\
&&~~~~~~~~
+ {{ 16 \sin^4 \theta R_\theta^8 h^2 } \over {r^{10} f^4 }} 
  \left \{ f_3 - {1 \over 2} ( h^0_{~0} + h^1_{~1} +h^r_{~r} ) \right \}
- {{ 10 \sin^2 \theta R_\theta^4 h } \over {r^{6} f^2 }} 
  h^r_{~r} = 0.
\label{philequation}
\end{eqnarray}
Our strategy is to disentangle the last three terms. 
For this purpose we have to use the dilaton gauge 
condition (\ref{dilaton-gauge}) and the linearized equations 
for $H, F_3, F_5$ in Eqs.(\ref{eq-hmnp})-(\ref{eq-fmnpqr}).
Because this is a nontrivial task for an 
arbitrary $\theta$, we only consider a simple and 
physically interesting case of $\theta \to {\pi \over 2}(B\to \infty)$. 
Here one choose the smeared D1-brane background
with $\bar H_{MNP}=\bar F_{MNPQR}=0$, which implies 
also that $h_3=f_5=0$.
Then Eq.(\ref{eq-hmnp}) with this gives us a constraint 
\begin{equation}
(\nabla_M \eta) \bar F^{MPQ} =0.
\label{Fconstraint}
\end{equation}
Since $\bar F^{MPQ} \ne 0$, Eq.(\ref{Fconstraint}) 
implies $\eta=0$.
This means that $\eta$ is a non-propagating mode in the 
smeared D1-brane background.
Hence the instability problem of the RR scalar arisen from the 
previous section is cured.
The remaining ones are the s-wave dilaton equation and $F_3$-equation
\begin{eqnarray}
&&\left \{
{ \partial^2 \over {\partial r^2}} +
{ 5 \over r} {\partial \over {\partial r}}
+ (\omega^2 - k_1^2 ) f - (k_2^2 + k_3^2 ) 
  + {{ 4 R_{\pi \over 2}^8 } \over {r^{10} f^2 } }
\right \} \hat \phi 
- {{ 10 R_{\pi \over 2}^8 h^r_{~r} } \over { r^6 f }} 
\nonumber \\
&&~~~~~~~~~~
+ {{ 16 R_{\pi \over 2}^8 } \over {r^{10} f^2}} 
  \left \{ ( f_3 - {1 \over 2} ( h^0_{~0} +h^1_{~1} +h^r_{~r} ) \right \} 
 =0,
\label{hatphi} \\
&&\nabla_M f^{MNP} - (\nabla_M h_Q^{~N} ) \bar F^{MQP} 
  + (\nabla_M h^P_{~Q} ) \bar F^{MQN} 
\nonumber \\
&&~~~~~~~~
  - (\nabla_M \hat h^M_{~Q} ) \bar F^{QNP} 
  - h^M_{~Q} (\nabla_M \bar F^{QNP} ) =0.
\label{fmnp}
\end{eqnarray}
Eq.(\ref{hatphi}) is derived from Eq. (\ref{philequation}) with 
$\phi^0 = h^{1/2} \hat \phi$ and $\theta = {\pi \over 2}$.
In order to decouple the last term in (\ref{hatphi}), 
we have to use both 
Eq.(\ref{fmnp}) and the dilaton gauge Eq.(\ref{dilaton-gauge}).
When $N=0, P=1$, solving Eq.(\ref{fmnp}) leads to\cite{Lee98PRD7361}
\begin{equation}
\partial_r \left ( f_3 - h^0_{~0} - h^1_{~1} \right )
  + \partial_0 h^0_{~r} + \partial_1 h^1_{~r} 
  +\left (
    { 5 \over r} + { f' \over f} 
   \right ) h^r_{~r} =0.
\label{rel-f3}
\end{equation}
Using the dilaton gauge, the last three terms turns out to be 
$\partial_r ( - h^r_{~r} + {1 \over 2} h )$.
Then Eq.(\ref{rel-f3}) gives us a crucial relation
\begin{equation}
 f_3 - { 1\over 2} (h^0_{~0}+h^1_{~1}+h^r_{~r})
  + { 1\over 2} (h^2_{~2}+h^3_{~3}+h^{\theta_i}_{~\theta_i}) =0.
\label{f3-const}
\end{equation}
We point out that the same relation (\ref{f3-const}) can be found if one 
uses the harmonic gauge (\ref{harmonic-gauge})\cite{Lee99JHEP10014}.
For simplicity, we can set $h^{\theta_i}_{~\theta_i}=0$ and $h^r_{~r}=0$.
Then the equation (\ref{hatphi}) leads to 
\begin{eqnarray}
&&\left \{
{ \partial^2 \over {\partial r^2}} +
{ 5 \over r} {\partial \over {\partial r}}
+ (\omega^2 - k_1^2 ) f - (k_2^2 + k_3^2 ) 
  + {{ 4 R_{\pi \over 2}^8 } \over {r^{10} f^2 } }
\right \} \hat \phi 
\nonumber \\
&&~~~~~~
- {{ 8 R_{\pi \over 2}^8 } \over { r^{10} f^2 }}(h^2_{~2}+h^3_{~3}) =0.
\label{hatphi-gauge}
\end{eqnarray}
If the last term is absent, Eq.(\ref{hatphi-gauge}) reduces to the RR 
scalar equation (\ref{hateta}) with $\theta={\pi\over 2}$.
It is easily proved that, considering (\ref{lin-chi}) and (\ref{lin-Phi}) only, 
one finds that the dilaton equation (\ref{eq-hmn}) leads to 
Eq.(\ref{hateta}).
Hence the presence of the last term is important
to distinguish the dilaton from the RR scalar.
Without $B$-field, the fixed scalar $\lambda$ is given by\cite{Gub98NPB393} 
\begin{equation}
\lambda = {{D-7} \over {2 \beta}} \Phi - { 1\over {2 \beta}} \log V,
\label{lambda}
\end{equation}
where $V$ is the world volume measured in $g_{MN}$.
This implies that a trace of gravitons($h^a_{~a}$) polarized parallel in the 
world volume plays a role of the fixed scalar.
With $B$-field, we may assume a relation between the dilaton and 
$h^a_{~a}$.
However, although we have a simple dilaton equation with the dilaton gauge, 
one cannot determine the relation between $\phi$ and 
$h^2_{~2} + h^3_{~3}$.
This is so because we have no further information for $h^2_{~2} + h^3_{~3}$.
Hence we also have to use the Einstein equation(\ref{eqRMN}) in the 
smeared D1-brane background of Appendix \ref{appendix}
Using Eq.(\ref{eqRMN}) and (\ref{eqPhi}), one obtains a scalar equation
\begin{equation}
R - 4 \left ( \nabla \Phi \right )^2 + 
  4 \nabla^2 \Phi = 0.
\label{R-equation}
\end{equation}
For an s-wave propagation, it is sufficient to consider 
Eq.(\ref{R-equation}) instead of Eq.(\ref{eqRMN}).
Its linearized equation takes the form 
\begin{eqnarray}
&& \bar G^{MN} \delta R_{MN}(h) -h^{MN} \bar R_{MN} 
  - 8 \left ( \nabla \bar \Phi \right ) \cdot \nabla \phi 
  - 4 \nabla_M \bar \Phi \nabla_N \bar \Phi h^{MN} 
\nonumber \\
&&~~~~~~
  - 4 h^{MN} \nabla_M \nabla_N \bar \Phi 
  - 4 \bar G^{MN} \delta \Gamma^P_{MN} \nabla_P \bar \Phi
  + 4 \nabla^2 \phi =0
\label{lin-R-eq}
\end{eqnarray}
with the Lichnerowitz operator\cite{Gre93PRL2837}
\begin{eqnarray}
\delta R_{MN}(h) &=& 
  - {1 \over 2} \nabla^2 h_{MN} 
  - {1\over 2} \nabla_M \nabla_N h 
  + {1 \over 2} \nabla^P \nabla_N h_{PM}
  + {1\over 2} \nabla^P \nabla_M h_{PN}
\label{lichnerowitz}\\
&=&
  - {1 \over 2} \nabla^2 h_{MN} 
  - \bar R_{Q(M} h^Q_{~N)} +\bar R_{PMQN} h^{PQ}
  + \nabla_{(M} \nabla_{|P|} \hat h^P_{~N)}.
\label{delR}
\end{eqnarray}
From Eq.(\ref{lichnerowitz}) we obtain
\begin{equation}
G^{MN} \delta R_{MN} = -\nabla^2 h + \nabla_P \nabla_N h^{PN} .
\label{tracedelR}
\end{equation}
The last term in (\ref{tracedelR}) with the dilaton gauge gives rise to 
a difficult relation for $h_{MN}$ to solve Eq. (\ref{lin-R-eq}).
Hence we would be better to use the harmonic gauge condition 
(\ref{harmonic-gauge}) to obtain
\begin{equation}
G^{MN} \delta R_{MN} = - {1 \over 2} \nabla^2 h,
\label{gdelr}
\end{equation}
which is also recovered from Eq.(\ref{delR}) with (\ref{harmonic-gauge}).

\section{Dilaton Propagation with harmonic gauge}
\label{harmonic-gauge-sec}
The equation (\ref{lin-R-eq}) leads to 
\begin{eqnarray}
&&\nabla^2(4 \phi - { h \over 2} ) 
  -{{4 f'} \over f^{3/2} } \phi' 
  -{1 \over f^{5/2} } ( 2 f f'' - f'^2 ) h^r_{~r} 
  - {{ 5 f'} \over { 2 r f^{3/2} }} h^{\theta_i}_{~\theta_i}
\nonumber \\
&&~~~~~~~~~~
  + {{ f'^2 ( h^0_{~0}+ h^1_{~1}- h^2_{~2}- h^3_{~3}+ h^r_{~r}- 
         h^{\theta_i}_{~\theta_i})} \over { 2 f^{5/2} }} =0.
\label{eq-phi-h}
\end{eqnarray}
Also the dilaton equation (\ref{eq-hmn}) takes the form
\begin{eqnarray}
&&\nabla^2\phi  
  -{{2 f'} \over f^{3/2} } \phi' 
  -{f'' \over {2 f^{3/2}} } h^r_{~r} 
  - {{ 2 f'} \over {  r f^{3/2} }} h^{\theta_i}_{~\theta_i}
\nonumber \\
&&~~~~~~~~~~
  + {{ f'^2 \{ h^0_{~0}+ h^1_{~1}- 5(h^2_{~2}+ h^3_{~3}+  
         h^{\theta_i}_{~\theta_i})\}} \over { 8 f^{5/2} }} =0.
\label{eq-dilaton}
\end{eqnarray}
Choosing Eq.(\ref{harmonic-gauge}) with $h=0$(transverse-traceless gauge), 
one has 
$h^0_{~0}+ h^1_{~1} = - (h^2_{~2}+ h^3_{~3})$ with 
$h^r_{~r}= h^{\theta_i}_{~\theta_i} =0$.
Then the above two equations become, respectively, 
\begin{equation}
\left \{
{\partial^2 \over {\partial r^2}} 
  +{ 5 \over r} {\partial \over {\partial r }} 
  + \tilde \omega^2 \left ( 1 + {\tilde R_{\pi\over 2}^4 \over r^4 } \right ) 
\right \} \hat \phi
  - {{ 4 R_{\pi\over 2}^8 ( h^2_{~2}+ h^3_{~3}) } \over {r^{10} f^2}} =0,
\label{eq-hatphi1}
\end{equation}
\begin{equation}
\left \{
{\partial^2 \over {\partial r^2}} 
  +{ 5 \over r} {\partial \over {\partial r }} 
  + \tilde \omega^2 \left ( 1 + {\tilde R_{\pi\over 2}^4 \over r^4 } \right ) 
  + {{ 4 R_{\pi\over 2}^8 } \over {r^{10} f^2}}
\right \} \hat \phi
  - {{ 12 R_{\pi\over 2}^8 ( h^2_{~2}+ h^3_{~3}) } \over {r^{10} f^2}} =0.
\label{eq-hatphi2}
\end{equation}
The two equations (\ref{eq-hatphi1}) and (\ref{eq-hatphi2}) 
should be the same.
Here we assume $h^2_{~2}+ h^3_{~3} = a \phi$.
Then one finds a relation
\begin{equation}
4 -12 a = -4 a ,
\label{rel-b}
\end{equation}
which gives us $a=1/2$.
Hence one obtains the correct dilaton equation as
\begin{equation}
\left \{
{\partial^2 \over {\partial r^2}} 
  +{ 5 \over r} {\partial \over {\partial r }} 
  +\tilde\omega^2 \left (1 + {{ \tilde R_{\pi\over 2}^4} \over r^4 } \right ) 
  - {{  2 R_{\pi\over 2}^8 } \over {r^{10} f^2}} 
\right \} \hat \phi = 0.
\label{cor-dil-eq}
\end{equation}
This can be approximated by using Eq. (\ref{rtilde}) as 
\begin{equation}
\left \{
{\partial^2 \over {\partial r^2}} 
  +{ 5 \over r} {\partial \over {\partial r }} 
  + \tilde \omega^2 \left ( 1 +{{\tilde R_{\pi\over 2}^4} \over r^4 }\right) 
  - {{ 2 \tilde R_{\pi\over 2}^8 } \over {r^{10} \tilde f^2}} 
\right \} \hat \phi \simeq 0
\label{app-dil-eq}
\end{equation}
for the leading-order calculation.
Finally, it remains to find an approximate solution to 
Eq.(\ref{app-dil-eq}) for low energies ($\tilde\omega \to 0$) and derive 
its absorption cross section.
We divide the space into three regions($I,II,III$) and then match 
solutions in them together.
In the near horizon region($I$) the equation takes the form
\begin{equation}
\left \{
{\partial^2 \over {\partial \rho^2}} 
  + { 5 \over \rho} {\partial \over { \partial \rho}} 
  + {{(\tilde \omega \tilde R_{\pi\over 2})^4} \over \rho^4 }
  - { 2 \over \rho^2}
\right \} \hat \phi_I (\rho) =0 ,
\label{near-eq}
\end{equation}
where $\rho = \tilde \omega r$.
Defining $\rho = {{(\tilde \omega \tilde R_{\pi\over 2})^2} \over z}$ and 
$\hat \phi_I(\rho) = z^{3/2} \hat{\hat \phi}_I$, this leads to
\begin{equation}
\left \{
{\partial^2 \over {\partial z^2}} 
  + 1 
  - { 23 \over {4 z^2}}
\right \} \hat{\hat \phi}_I (z) =0,
\label{near-bessel}
\end{equation}
which is nothing but the standard Bessel equation for 
$\hat{\hat\phi}_I(z) = H_{\sqrt{6}}(z)$.
The above equation can be interpreted as the Schr\"odinger-like 
equation with the energy $E=1$ which is valid for large $z$(in the 
near horizon of $r\to 0$).
The solution is given by
\begin{equation}
\hat \phi_I(z) = z^2 H_{\sqrt{6}}(z).
\label{near-sol}
\end{equation}
In the intermediate zone($II$), the $\tilde \omega$-term can be ignored.
Thus one finds the solution
\begin{equation}
\hat \phi_{II}(\rho) = C 
\left \{
{{\rho^4} \over {(\tilde \omega \tilde R_{\pi\over 2} )^4 
        + \rho^4 }}
\right \}^{ { 1\over 2} \sqrt{ 3 \over 2} - {1 \over 2} } .
\label{inter-sol}
\end{equation}
In the far infinity region($III$) we have the equation
\begin{equation}
\left (
{ \partial^2 \over { \partial \rho^2}} 
  + { 5 \over \rho} { d \over { d \rho} } 
  + \tilde \omega^2 
\right ) \hat \phi_{III}(\rho) =0.
\label{far-eq}
\end{equation}
Its solution is given by
\begin{equation}
\hat \phi_{III}(\rho) = D {{ J_2(\rho) } \over \rho^2 }.
\label{far-sol}
\end{equation}
Matching $III$ to $II$ leads to
\begin{equation}
D = 8C.
\label{match-23}
\end{equation}
Also matching $I$ to $II$ gives
\begin{equation}
C = {{ 2^{\sqrt{6}}} \over \pi} \Gamma(\sqrt{6}) 
   (\tilde \omega \tilde R_{\pi \over 2} ) ^{2 -\sqrt{6} } .
\label{match-12}
\end{equation}
Considering the ratio of the flux at the horizon($r=0$) to the 
incoming flux at infinity leads to the 
absorption probability as
\begin{equation}
P_\phi = { 4 \over {| D |^2} } (\tilde \omega \tilde R_{\pi \over 2} ) ^8
  = { 1\over 16} {\pi^2 \over { 2^{2\sqrt{6}} }}
    {{(\tilde \omega \tilde R_{\pi \over 2} ) ^{2 \sqrt{6} +4} } \over 
             {\Gamma(\sqrt{6} )^2}}.
\label{abs-prob}
\end{equation}
Finally, we obtain the s-wave absorption cross section of the dilaton 
in the limit of $B\to\infty$ as
\begin{eqnarray}
\sigma_0^\phi \Big \vert_{B\to\infty} &=& 
  {{2^5 \pi^2} \over \omega^5} P_\phi =
  {\pi^4 \over {2^{2 \sqrt{6}-1} \Gamma(\sqrt{6})^2 } }
  {{(\tilde \omega \tilde R_{\pi \over 2} ) ^{4 + 2 \sqrt{6} } } \over 
           \omega^5 }
\nonumber \\
  &\simeq &
   { \pi^4 \over {2^{2 \sqrt{6}-1} \Gamma(\sqrt{6})^2 } }
   {{(\tilde \omega \tilde R_{\pi \over 2} ) ^{8.9} } \over \omega^5} .
\label{sigma-phi1}
\end{eqnarray}

\section{Discussions}
\label{discussions}
First we discuss the propagation of fields in the 
smeared D1-brane background of $B\to\infty$ limit.
We have shown that, considering $h=h^r_{~r} = h^{\theta_i}_{~\theta_i} =0$, 
the dilaton $\phi$, 
$f_3$, and $h^2_{~2} + h^3_{~3}= - h^1_{~1} -h^2_{~2}=\phi/2$ are
physically propagating modes whereas
the RR scalar $\eta$, $h_3$, and $f_5$, are non-propagating modes.
Interestingly, it turns out that the absorption cross section of 
the dilaton in the limit of $B\to\infty$ 
is given by the replacement of $R_0 \to \tilde R_{\pi\over2}$ and
$8 \to 8.9$ in Eqs.(\ref{sigmaB0l0}) and (\ref{sigma-phi1}).
The $R_0 \to \tilde R_{\pi\over 2} ( 8 \to 8.9)$ arise from 
the presence of $B$-field(the couplings).

The RR scalar has a negative potential as shown in Fig.\ref{rr-scalar}.
This induces an instability of the near horizon 
in the smeared D1-brane background.
Fortunately, $H$-equation (\ref{eq-hmnp})
requires that this mode should not be a propagating one.

For a general analysis, let us consider the following equation 
with the parameter $s$ upon the diagonalization:
\begin{equation}
\left \{
{{\partial^2} \over { \partial r^2}} 
  + { 5 \over r} {\partial \over {\partial r}}
  + \tilde \omega^2 \left ( 1 + {{\tilde R_{\pi\over 2}^4 }\over r^4 } \right )
  - {{ s \tilde R_{\pi\over 2}^8} \over { r^{10} \tilde f^2 }} 
\right \} \psi_s = 0.
\label{eq-psis}
\end{equation}
For $s > -4$, one obtains its absorption cross section as
\begin{equation}
\sigma_0^{\psi_s} \Big\vert_{B\to\infty} =
  {\pi^4 \over { 2^{2 \sqrt{4+s} -1 } \Gamma(\sqrt{4+s})^2}}
  {{(\tilde \omega \tilde R_{\pi\over2} )^{2 \sqrt{4+s} +4 } }
          \over \omega^5}  .
\label{abs-psis}
\end{equation}
For the case of $s=32$ case, one finds an interesting cross section
\begin{equation}
\sigma_0^{\psi_{s=32}} \Big\vert_{B\to\infty} =
  {\pi^4 \over { 2^{14} \cdot 15} } 
  {{(\tilde \omega \tilde R_{\pi\over2} )^{16} }
          \over \omega^5}  ,
\label{abs-psi32}
\end{equation}
which is the same order as in $h^a_{~a}$ in the absence of 
$B$-field and $k^2=0$\cite{Gub98NPB393}
\begin{equation}
\sigma_0^{h^a_{~a}} \Big\vert_{B=0} =
  {\pi^4 \over { 2^{17} \cdot 3^4} } 
  {{(\omega R_0 )^{16} }
          \over \omega^5}  .
\label{abs-haa}
\end{equation}
Hence we expect that the new scalars in 
the limit of $B\to\infty$ may take a value of $ -4 \le s \le 32$.
Especially, the RR scalar with $s=-4$ is not allowed for matching 
procedure and thus it cannot be a propagating mode.
And the dilaton has $s=2$ and its absorption cross section is given by
(\ref{sigma-phi1}).
For $s=0$ case(minimally coupled scalar), one can recover 
(\ref{sigmaB}) with $\theta={\pi\over2}$ from (\ref{abs-psis}).

In conclusion, the $B\to\infty (\theta\to{\pi\over 2})$ limit is a 
delicate issue.
Here for the calculation of the absorption cross section, we take 
only the limit of $\alpha' \to 0$.
In this case one finds ${\bar H} \propto \alpha'$, 
${\bar F}_3 \propto 1/\alpha'$, 
${\bar F}_5 \propto {\rm finite}$.
It is known that ${\bar F}_5$ counts the rank of the 
noncommutative group.
However, ${\bar F}_5 \ne 0$ and ${\bar H} \ne 0$ leads to the 
complicated coupled equations.
Solving these coupled equations is a formidable task.
We remind the reader that the fluxes of ${\bar F}_5$ and 
${\bar H}$ can be neglected 
in comparison with that of ${\bar F}_3$.
Hence we choose the simple smeared D1-brane background by setting 
${\bar F}_5 = {\bar H} =0$.
At the first sight, this action seems to be eliminating any connection 
to noncommutative geometry.
However, although we do not count the fluxes of $F_5$ and $H$ thoroughly, 
we still give the effect of the noncommutativity on the absorption 
cross section through $\tilde R_{\pi\over 2}^4 \gg R_0^4$ in 
$ f = 1 + \tilde R_{\pi\over 2}^4 / r^4$ and the important coupling of $F_3$.
If ${\bar H} \ne 0$ and ${\bar F}_5 \ne 0$, 
we expect that there will be a change in $s$ : 
$R_0\to \tilde R_{\pi\over 2}$, $8\to s ( 2 \le s < 32)$.
This is so because the coupling scheme will change ``$s$''.

Finally we comment the ways to account the noncommutative effect 
on the cross section of the dilaton on the supergravity side.
These are $R_{\pi\over 2}^4 \gg R_0^4$, $k^2 \ne 0$, and 
the couplings to all other fields.
Here we include the expansion of the parameter $R_{\pi\over 2}^4(\gg R_0^4)$, 
the presence of momenta along the world volume 
directions($k_2, k_3$) to detect 
$B_{23}$-field, and the coupling of $F_3$ with $H = F_5 =0$.
Analysis for an arbitrary $\theta(B)$ remains unexplored.

\section*{Acknowledgement}
Authors would like to thank Sangmin Lee for helpful discussion.
This work was supported by the Brain Korea 21
Program, Ministry of Education, Project No. D-0025.
\appendix
\section{The smeared D1-brane solution}
\label{appendix}
In the case of $F_5=H_3=\chi$, the string frame action takes the form
\begin{equation}
S_{10}^{\rm SD1} = {1 \over {2 \kappa_{10}^2 }} 
\int d^{10} x \sqrt{-G~} \left [
e^{-2 \Phi} \left \{ R + 4 \left ( \nabla \Phi \right )^2 
- {1 \over 12} F_3^2 \right \}
\right ],
\label{saction}
\end{equation}
which leads to the equations of motion
\begin{eqnarray}
&&R_{MN} = -2 \nabla_M \nabla_N \Phi
 + { 1\over 4} e^{2\Phi} F_{MPQ} F_{N}^{~~PQ}
 - {1\over 24} e^{2 \Phi} F_3^2 G_{MN}, 
\label{eqRMN} \\
&& \nabla^2 \Phi - 2 \left ( \nabla \Phi \right )^2 
  - {1 \over 12} e^{2 \Phi} F^2 = 0,
\label{eqPhi} \\
&& \nabla_M F_3^{MPQ} = 0.
\label{eqF3} 
\end{eqnarray}
The smeared D1-brane solution is given by 
\begin{eqnarray}
&&ds_{\rm SD1}^2 = f^{-{1\over 2}} \left \{
-dx_0^2 + dx_1^2 + f \left ( dx_2^2 + dx_3^2 \right ) 
\right \}
+ f^{1\over 2} \left ( dr^2 + r^2 d\Omega_5^2 \right ) ,
\nonumber \\
&& f = 1 + { C \over r^4}, ~~ e^{2 \bar \Phi} = g^2 f, 
~~ {\bar F}_{01r} = {1 \over g} \partial_r(f^{-1} ).
\label{smeared}
\end{eqnarray}
Here ``C'' is an arbitrary constant, but in order to make connection 
to the noncommutative geometry we have to choose 
$C=R_{\pi\over 2}^4 = 4 \pi \tilde g \tilde b \tilde N \gg R_0^4$.

\end{document}